\documentclass[10pt]{article}

\usepackage{fullpage}
\usepackage{setspace}
\usepackage{parskip}
\usepackage{titlesec}
\usepackage[section]{placeins}
\usepackage{xcolor}
\usepackage{breakcites}
\usepackage{lineno}
\usepackage{hyphenat}

\PassOptionsToPackage{hyphens}{url}
\usepackage[colorlinks = true,
            linkcolor = blue,
            urlcolor  = blue,
            citecolor = blue,
            anchorcolor = blue]{hyperref}
\usepackage{etoolbox}

\usepackage{natbib}

\renewenvironment{abstract}
  {{\bfseries\noindent{\abstractname}\par\nobreak}\footnotesize}
  {\bigskip}

\titlespacing{\section}{0pt}{*3}{*1}
\titlespacing{\subsection}{0pt}{*2}{*0.5}
\titlespacing{\subsubsection}{0pt}{*1.5}{0pt}

\usepackage{authblk}

\usepackage{graphicx}
\usepackage[space]{grffile}
\usepackage{latexsym}
\usepackage{textcomp}
\usepackage{longtable}
\usepackage{tabulary}
\usepackage{booktabs,array,multirow}
\usepackage{amsfonts,amsmath,amssymb}

\usepackage{bm}
\usepackage{enumitem}

\providecommand\citet{\cite}
\providecommand\citep{\cite}

\newif\iflatexml\latexmlfalse

\AtBeginDocument{\DeclareGraphicsExtensions{.pdf,.PDF,.eps,.EPS,.png,.PNG,.tif,.TIF,.jpg,.JPG,.jpeg,.JPEG}}

\usepackage[utf8]{inputenc}
\usepackage[english]{babel}
\usepackage{float}
\usepackage[margin=1.5in]{geometry}

\newcommand\dd{\mathrm{d}}

\usepackage{algorithm}
\usepackage{algcompatible}

\begin{document}

\title{Sampling Constrained Continuous Probability Distributions: A Review}

\author[1]{Shiwei Lan}%
\affil[1]{School of Mathematical and Statistical Sciences, Arizona State University}%
\author[*2]{Lulu Kang}
\affil[2]{Department of Applied Mathematics, Illinois Institute of Technology}

\vspace{-1em}

\date{}

\begingroup
\let\center\flushleft
\let\endcenter\endflushleft
\maketitle
\endgroup

\selectlanguage{english}
\begin{abstract}
The problem of sampling constrained continuous distributions has frequently appeared in many machine/statistical learning models. 
Many Monte Carlo Markov Chain (MCMC) sampling methods have been adapted to handle different types of constraints on the random variables. 
Among these methods, Hamilton Monte Carlo (HMC) and the related approaches have shown significant advantages in terms of computational efficiency compared to other counterparts. 
In this article, we first review HMC and some extended sampling methods, and then we concretely explain three constrained HMC-based sampling methods, reflection, reformulation, and spherical HMC. 
For illustration, we apply these methods to solve three well-known constrained sampling problems, truncated multivariate normal distributions, Bayesian regularized regression, and nonparametric density estimation. 
In this review, we also connect constrained sampling with another similar problem in the statistical design of experiments of constrained design space. 

\end{abstract}%

\section{Introduction}\label{sec:intro}

In many machine learning applications, it is necessary to sample from distributions with various types of constraints. 
For example, the truncated multivariate normal distribution can be difficult to sample from, especially for high-dimensional cases.  
Even using the leave-one-out type of Gibbs sampling scheme \citep{held2006bayesian, kang2021bayesian}, the algorithms can still be computationally costly. 
Another common example is the regression model with norm constraint on the parameters, $\|\bm \beta\|_q\leq C$, such as Lasso ($l_1$ norm, $q=1$) \citep{tibshirani1996regression} or bridge estimator ($l_q$ norm, $q\geq 0$) \citep{frank1993statistical,fu1998penalized}. 
Other examples include copula models, latent Dirichlet allocation, covariance matrix estimation, and nonparametric density function estimation. 
Often, the resulting models are intractable, and thus sampling from these constrained distributions is a challenging task \citep{pneal08,sherlock09, pneal12, brubaker12, pakman14}. 
In this article, we give an overview of the statistical sampling methods for constrained distributions. 
Specifically, we focus on the distributions of continuous variables and provide a more detailed explanation of methods based on Hamiltonian Monte Carlo (HMC) and some related approaches. 
Sampling constrained discrete distributions \citep{Jessen_1970,jacob2021gibbs,chewi2022rejection} is not included in this review due to the different nature from continuous distribution. 

There are various types of boundary constraints on the parameters of many statistical models, such as positive requirement, linear (summation) constraint, upper bound on a vector norm of the parameters, etc. 
Sometimes, the constraints can be considered as certain general manifolds, e.g. sphere, positive definite matrices, and Stiefel manifold. 
Many sampling approaches have been introduced to tackle one or several kinds of constraints. 
Based on the nature of these methods, we roughly categorize them into three groups. 
\begin{enumerate}
\item {\bf Rejection Type}. These methods simply discard samples that violate the constraints or keep trying until the proposed sample satisfies the constraint. 
Most Markov Chain Monte Carlo (MCMC) or other statistical sampling algorithms can be easily modified to achieve this goal. 
For example, \cite{LANG_2007} proposed a rejection based sequential Monte Carlo for Bayesian estimation of constrained dynamic system, and \cite{Li_2015} also developed methods based on rejection for truncated multivariate normal and student-$t$ distributions subject to linear inequality constraints.
Since these approaches do not directly address the constraints, they can be computationally inefficient for complicated constraints and high-dimensional problems. 

\item {\bf Reflection Type}. These approaches consider the boundary of the constrained domain as an (energy) wall and make a reflection (hit-and-bounce) for the sampler to move inward the constrained domain whenever it hits the (energy) wall. 
For example, \cite{neal11} suggested modifying the standard HMC by setting the potential energy to infinity for parameter values that violate the constraints. 
Following this idea, \cite{pakman14} proposed an exact HMC for truncated multivariate Gaussian distributions. 
\cite{betancourt2011nested} and \cite{olander_2020} applied such HMC-based reflection idea to nested sampling which requires likelihood-restricted prior sampling \citep{skilling2006nested}. 

\item {\bf Reformulation Type}. These techniques transform the constrained sampling problem or the constrained domain into something easier to work with. 
Motivated by the constrained optimization methods, \cite{brubaker12} proposed a family of HMC-based MCMC methods, which incorporated the constraint on parameters $c(\bm \theta)=0$ using Lagrange multipliers. 
\cite{ahn2021efficient} recently derived another optimization motivated algorithm using mirror-Langevin dynamics.
In many cases, distributions with constraints in $\mathbb{R}^d$ can be transformed into distributions on manifolds. 
Some HMC-related algorithms have been created to sample distributions on manifolds. 
For example, \cite{Kook_2022} used Riemannian manifold HMC instead of the original HMC with a Lagrange multiplier. 
\cite{byrne13} showed how HMC methods can be designed for and applied to the distribution defined on manifolds embedded in Euclidean space by the explicit forms for geodesics if they exist. 
In particular, motivated by \cite{byrne13}, Spherical HMC \citep{lan14b,Lan2016b} does not require any manifold embedding. 
It focuses on constraints that can be transformed into vector norms and eventually mapped onto a hyper-sphere.
Spherical HMC can be viewed as a more efficient special case of \cite{brubaker12} and \cite{Kook_2022}.
A related work SPInS \citep{Chaudhry_2021} maps a hyper-ball containing the constrained domain inside out so that sampler can be defined on a larger unbounded space. 
\end{enumerate}

In general, algorithms of the rejection type tend to be inefficient because frequent rejected attempts cause a significant waste of computation. 
The reflection type is intuitive but its efficiency usually depends on the specific constraints. 
Certain constrained domains may need too many reflections in high dimensional space and hence the excessive computational time. 
The reformulation type is more intrinsic and sophisticated in incorporating the constraints in the step of proposing new samples. 
More importantly, they usually can be easily scaled to large dimensions and handle more varieties of constraints. 
Therefore, we are going to focus on the reflection and reformulation types of methods in the following review. 

In recent years, Optimal Transport (OT) and the more general variational inference methods have gained much attention from the machine learning community and have been adapted to many statistical and machine learning models.
Although in theory variational inference methods, including OT, can deal with distributions with any compact support regions, in numerical implementation it is much more challenging when the support region is not the entire $\mathbb{R}^d$ with $d$ being the dimension of the variables. 
\cite{das2020optimal} addressed the issue of state-dependent nonlinear equality-constrained state estimation using Bayesian filtering based on OT.
\cite{ahn2021efficient} proposed using the mirror-Langevin algorithm, which is a discretization of the mirror-Langevin diffusion, for constrained sampling.

Constrained sampling is naturally connected with numerical integration in constrained domains. 
Quasi-Monte Carlo methods deal with numerical integration over box constrained domains \citep{mcbook,leobacher2014introduction}. 
Other quadrature methods have to be modified for non-box constraints \citep{olshanskii2016numerical,gessner2020integrals,legrain2021non,saye2022high}.
Constrained sampling is also related to the statistical design of experiments.
\cite{draguljic2012noncollapsing}, \cite{pratola2017design}, and \cite{huang2021constrained} proposed different methods to generate space-filling designs, which essentially approximate the uniform distribution, in various irregular-constrained domains. 
 \cite{kang2019stochastic} developed algorithms to generate different optimal designs, including a distance-criterion-based space-filling design, in complicated design regions. 

 In the remaining article, we first review the necessary background on HMC and its related methods in Section \ref{sec:hlmc}.
 Next, we review in detail three reflection and reformulation approaches in Section \ref{sec:methods}.
Three constrained sampling problems are illustrated in Section \ref{sec:examples}, including truncated multivariate Gaussian distributions, Bayesian regularized regression, and non-parametric density estimation. 
In Section \ref{sec:doe}, we connect constrained sampling with the constrained design of experiments and review some existing methods for constructing different types of constrained designs. 
The article concludes in Section \ref{sec:end}. 

\section{Hamiltonian Monte Carlo and Extensions}\label{sec:hlmc}

MCMC can be inherently inefficient due to its random walk nature. 
Different from the Gibbs sampler and Metropolis algorithms, HMC simulates the Hamiltonian dynamics to propose new states and reduce the local random walk behavior, and thus moves more rapidly towards the target distribution. 
The proposed states of HMC are significantly distant from the current states and yet still have a high acceptance probability. 
\cite{neal11} recalled the origin and history of HMC and detailed the HMC method and its appealing properties. 
Here we briefly review the HMC algorithm and its manifold extensions such as the Riemannian/Lagrangian Monte Carlos, geodesic Monte Carlo, etc. 

\subsection{Hamiltonian Monte Carlo}\label{subsec:hmc}

We begin with an introduction to the Hamiltonian dynamics, which is the basis of HMC.
It consists of a $d$-dimensional vector $\bm \theta$, called \emph{position} state, and a $d$-dimensional vector $\bm \phi$, called \emph{momentum} state. 
For illustration, \cite{neal11} used a simple physical example, the dynamics of a frictionless puck that slides over a surface of varying height. 
The \emph{potential energy} of the puck, denoted by $U(\bm \theta)$, is proportional to the height given position $\bm \theta$. 
The \emph{kinetic energy} of the puck, denoted by $K(\bm \phi)$, is equal to $|\bm \phi|^2/m$, with $m$ the \emph{mass} of the puck. 
The total energy of the dynamic, called \emph{Hamiltonian function} and denoted by $H(\bm \theta, \bm \phi)$, is the sum of the potential energy and kinetic energy of the puck, i.e., 
\begin{equation}\label{eq:hamiltonian}
H(\bm \theta, \bm \phi)=U(\bm \theta)+K(\bm \phi). 
\end{equation}
As the puck slides over the surface, the potential energy increases as it moves over a rising slope and the kinetic energy decreases as the velocity of the puck, $\bm \phi/m$, decreases. 
The two energies changes in opposite directions when the puck moves over a descending slope. 
Due to frictionless assumption, the total energy remains the same as the initial state of the system. 
The system of $(\bm \theta, \bm \phi)$ evolves following the \emph{Hamilton's equations}
\begin{align}\label{eq:he-1}
\dot{\bm \theta}=\frac{\dd \bm \theta}{\dd t} &= \frac{\partial H(\bm \theta, \bm \phi)}{\partial \bm \phi}=\nabla_{\bm \phi}K(\bm \phi),\\\label{eq:he-2}
\dot{\bm \phi}=\frac{\dd \bm \phi}{\dd t} &=- \frac{\partial H(\bm \theta, \bm \phi)}{\partial \bm \theta}=-\nabla_{\bm \theta}U(\bm \theta).
\end{align}

HMC applies the Hamiltonian dynamics to MCMC sampling. 
The position vector $\bm \theta$ is the random variable of interest. 
The potential energy is $U(\bm \theta)=-\log p(\bm \theta)$, where $p(\bm \theta)$ is the density function of the target distribution to be sampled from. 
In the Bayesian framework, $p(\bm \theta)$ is the posterior distribution $p(\bm \theta|\bm y)$, where $\bm y$ represents data and $\bm \theta$ the unknown parameters. 
The momentum vector $\bm \phi$ can be considered as an auxiliary random variable that is usually assumed to follow a multivariate normal distribution $\mathcal{N}({\bf 0}, \bm M)$, where $\bm M$ is a user-specified covariance matrix (often chosen to be identity matrix $\bm I$), also known as a \emph{mass matrix}. 
Thus the kinetic energy becomes $K(\bm \phi)=-\log N(\bm \phi|{\bf 0}, \bm M)=\bm \phi^\top \bm M^{-1}\bm \phi/2+\text{constant}$. 
To numerically solve \eqref{eq:he-1} and \eqref{eq:he-2} with the above potential and kinetic energies, the \emph{leapfrog} method \citep{verlet67} is commonly used to approximate the Hamilton's equations by discretizing time. 
The HMC uses the leapfrog method to simulate the Hamiltonian dynamics for some time horizon $\tau=L\epsilon$ to propose new samples that are further accepted or rejected according to certain probability as the next state.
Algorithm \ref{alg:HMC} shows the HMC procedure.

\begin{algorithm}[t]
\caption{Hamiltonian Monte Carlo (HMC)}
\label{alg:HMC}
\begin{algorithmic}[1]
\STATE Initialize $\bm \theta^{(0)}$ at current $\bm \theta_{t-1}$, and randomly sample $\bm \phi^{(0)} \sim \mathcal{N}({\bf 0}, \bm M)$.
\FOR{$\ell=0$ to $L-1$}
\STATE Update $\bm \phi$ by a half-step of $\epsilon$: $\bm \phi^{(\ell+\frac{1}{2})} = \bm \phi^{(\ell)}+\frac{1}{2}\epsilon \nabla_{\bm \theta}\log p(\bm \theta^{(\ell)})$.
\STATE Update $\bm \theta$ by a full-step of $\epsilon$: $\bm \theta^{(\ell+1)} = \bm \theta^{(\ell)} + \epsilon \bm M^{-1}\bm \phi^{(\ell+\frac{1}{2})}$.
\STATE Update $\bm \phi$  by another half-step of $\epsilon$: $\bm \phi^{(\ell+1)} = \bm \phi^{(\ell+\frac{1}{2})}+\frac{1}{2}\epsilon \nabla_{\bm \theta}\log p(\bm \theta^{(\ell+1)})$.
\ENDFOR
\STATE Set $(\bm \theta^*, \bm \phi^*) = (\bm \theta^{(L)}, \bm \phi^{(L)})$ and compute the accept rate $r=\frac{p(\bm \theta^*)N(\bm \phi^*|{\bf 0}, \bm M)}{p(\bm \theta_{t-1})N(\bm \phi_{t-1}|{\bf 0}, \bm M)}$.
\STATE Set $\bm \theta_t=\bm \theta^*$ with probability $\min(r, 1)$ and $\bm \theta_{t}=\bm \theta_{t-1}$ otherwise.
\end{algorithmic}
\end{algorithm}


Following either the Hamilton's equations \eqref{eq:he-1} and \eqref{eq:he-2} or the leapfrog procedure, the intuition of HMC is straightforward as explained in \cite{gelman2013bayesian}. 
When the current value of $\bm \theta$ is at a flat region of $p(\bm \theta)$, similar to the situation when the puck is on a flat surface, the velocity of the puck, and thus its momentum $\bm \phi$, becomes close to a constant. 
Therefore, the position $\bm \theta$ would move at a constant speed exploring the flat region. 
If the position $\bm \theta$ moves to a region with decreasing density of $p(\bm \theta)$, which is not favorable, then $\nabla_{\bm \theta}\log p(\bm \theta)$ is negative, and thus the momentum $\bm \phi$ would decrease in the direction of movement. 
Next, the position $\bm \theta$ would move in this unfavorable direction with a reduced velocity. 
The trends reverse if $\bm \theta$ value moves to a region with an increasing density of $p(\bm \theta)$. 
The system possesses three important properties for the proof of ergodicity:
(i) time-reversibility (going from the end of a trajectory with the reversed momentum takes the sampler back to the starting point); (ii) volume-preservation (volume moving along the flow $T_t: (\bm\theta, \bm\phi)\mapsto (\bm\theta^*, \bm\phi^*)$ does not change); and (iii) energy conservation (approximately under discretized system).
The exact proof of convergence of HMC can be found in \cite{NEAL1994194} and \cite{neal11}. 
The choice of $\epsilon$ should be sufficiently small so that the acceptance rate is high but not too small so that the computation is still efficient. 
The length of trajectory is also a crucial parameter for HMC and it can be varied from 20 to 1000, depending on the complexity and dimension of the problem. 
Trial and error can be used for setting both $\epsilon$ and $L$. 
More discussion can be found in \cite{neal11}. 

\subsection{Riemannian and Lagrangian Monte Carlo}\label{subsec:lmc}

\cite{girolami11} extended HMC to Riemannian HMC (RHMC) by defining the Hamiltonian dynamics on a Riemannian manifold of distributions. 
Compared to HMC, RHMC can exploit fully the geometric properties of the parameter space of $\bm \theta$ using a position-specific mass matrix, i.e., $\bm M=\bm G(\bm \theta)$, where $\bm G(\bm \theta)$ is usually set as the Fisher information matrix of $p(\bm \theta)$. 
The distribution of momentum vector $\bm \phi$ is $\mathcal{N}({\bf 0}, \bm G(\bm \theta))$, which is no longer independent of $\bm \theta$. 
The Hamiltonian function \eqref{eq:hamiltonian} becomes 
\begin{equation}\label{eq:mhamiltonian}
    H(\bm \theta, \bm \phi) = \psi(\bm \theta) + K(\bm\theta, \bm\phi) = -\log p(\bm \theta)+ \frac{1}{2}\log \det(\bm G(\bm \theta)) + \frac{1}{2}\bm \phi^\top \bm G(\bm \theta)^{-1}\bm \phi.
\end{equation}
where $\psi(\bm \theta)=-\log p(\bm \theta)+\frac{1}{2}\log \det(G(\bm \theta))$, and $K(\bm\theta, \bm\phi)=\frac{1}{2}\bm \phi^\top \bm G(\bm \theta)^{-1}\bm \phi$.
Due to dependence of $\bm \phi$ on $\bm \theta$, the dynamics of $\bm \theta$ and $\bm \phi$ become non-separable. The previous version of the leapfrog is not applicable. 
Instead, the \emph{generalized leapfrog} \citep{10.1093/imanum/6.4.381} is used, which is an implicit scheme of fixed-point iterations.  

To avoid the time-consuming iterations, \cite{lan14a} introduced the \emph{Lagrangian Monte Carlo} (LMC).
It uses a variable transformation approach that changes the Hamiltonian dynamics to Lagrangian dynamics. 
Specifically, let $\bm v = \bm G(\bm \theta)^{-1}\bm \phi$ and its distribution is $\bm v\sim \mathcal{N}({\bf 0}, \bm G(\bm \theta)^{-1})$.
As $\bm v$ is the momentum divided by mass, it can be considered as velocity intuitively.
The original Hamilton's equations in HMC \eqref{eq:he-1} and \eqref{eq:he-2} become the following Lagrangian dynamics (a.k.a. Euler-Lagrange equation):
\begin{align}\label{eq:le-1}
\frac{\dd \bm \theta}{\dd t}&=\bm v,\\ \label{eq:le-2}
\frac{\dd \bm v}{\dd t} &= -\bm \eta(\bm \theta, \bm v)-\bm G(\bm \theta)^{-1}\nabla_{\bm \theta}\psi(\bm \theta), 
\end{align}
where $\bm \eta(\bm \theta,\bm v)$ is a vector whose $k$th element is $\bm v^\top \bm \Gamma^k(\bm \theta)\bm v$.
Here $\bm \Gamma^k_{i,j}(\bm \theta):= \frac{1}{2}\sum_{l}g^{k,l}(\partial_{i} g_{l,j}+\partial_j g_{i,l}-\partial_l g_{i,j})$ are the Christoffel symbols, where $g_{i,j}=[\bm G(\bm \theta)]_{i,j}$ and $g^{i,j}=[\bm G(\bm \theta)^{-1}]_{i,j}$ and $\partial_i$ means partial derivative with respect to $\theta_i$.
Based on the Lagrangian dynamics, \cite{lan14a} proposed an explicit integrator, which is time reversible but not volume preserving.
This is different from the HMC and RHMC.
However, one can adjust the acceptance probability with the Jacobian determinant to satisfy the detailed balance condition.
The LMC algorithm in \cite{lan14a} is shown to be computationally more stable and efficient than RHMC.

\cite{byrne13} developed these manifold HMC algorithms for a class of problems where the geodesic equation (the system \eqref{eq:le-1} and \eqref{eq:le-2} without $\bm G(\bm \theta)^{-1}\nabla_{\bm \theta}\psi(\bm \theta)$ term) can be analytically solved.
\cite{lan14b,Lan2016b} proposed a specific geodesic MC on a hyper-sphere (whose geodesic is a big circle) and applied it to sample from distributions with constraints defined by vector norm.

\section{Constrained Sampling based on HMC}\label{sec:methods}

In this section, we explain three different constrained sampling methods which are adapted from the original HMC. 
They are \emph{Wall HMC}, \emph{Constrained HMC}, and \emph{Sphere HMC}, which are among the most representative ones in the reflection and reformulation types of algorithms. 

\subsection{Constrained HMC by Reflection}
\cite{neal11} discussed a method of handling the constraint $c(\bm\theta)\geq 0$ by modifying the original potential energy $U(\bm\theta)$ to create a ``soft" wall:
\begin{equation}
U_r(\bm\theta) = U(\bm\theta) + w(\bm\theta), \quad
w(\bm\theta) = 
\begin{cases}
    0, & \;\text{if}\; c(\bm\theta)\geq 0, \\
    r^{r+1}|c(\bm\theta)|^r, & \text{else}.
\end{cases}
\end{equation}
Such constraint becomes a ``hard" wall with infinite barrier
\begin{equation}
\tilde U(\bm\theta) = \lim_{r\to\infty} U_r(\bm\theta) =
\begin{cases}
    U(\bm\theta), & \;\text{if}\; c(\bm\theta)\geq 0, \\
    \infty, & \text{else}.
\end{cases}
\end{equation}

Suppose the sampler just hits the wall, i.e. $c(\bm\theta)<0$.
According to \eqref{eq:he-2} with $U_r$, we have the momentum $\bm\phi$ updated as
\begin{equation}
\bm\phi_{t+1} = \bm\phi_t -\nabla_{\bm \theta}U(\bm \theta) \Delta t - r^{r+2} \frac{|c(\bm \theta)|^r}{c(\bm\theta)} \Delta t \nabla_{\bm \theta}c(\bm \theta)
\end{equation}
We can choose the time step $\Delta t = C(r) \to 0$ as $r\to\infty$ such that we have the following perfect reflection for updating momentum \citep{betancourt2011nested}
\begin{equation}\label{eq:bounce}
\bm\phi_{t+1} = \bm\phi_t - 2 \langle \bm\phi_t, \bm n\rangle \bm n, \quad \bm n= \nabla_{\bm \theta}c(\bm \theta)/\Vert \nabla_{\bm \theta}c(\bm \theta)\Vert. 
\end{equation}

Such reflection based HMC (named as ``Wall HMC'' in \cite{lan14b}) proceeds with $\tilde U(\bm\theta)$ replacing $U(\bm\theta)$ in Algorithm \ref{alg:HMC} where the half-step momentum updates (lines 3 and 5) are replaced by the above reflection \eqref{eq:bounce} once hitting the wall (constraint violated). 
\cite{betancourt2011nested} and \cite{olander_2020} successfully applied this algorithm to nest sampling. 
\cite{pakman14} developed a more sophisticated exact HMC algorithm for truncated multivariate Gaussian distributions based on a similar idea of reflection.

\subsection{Constrained HMC by Reformulation}
\cite{brubaker12} considered HMC for a general constraint $c(\bm\theta)=0$ that defines a connected, differentiable submanifold of $\mathbb R^d$, denoted as $\mathcal M=\{\bm\theta\in \mathbb R^d| c(\bm\theta)=0 \}$.
The constraint determines the \emph{tangent bundle} of $\mathcal M$, $\mathcal{TM}=\{(\bm\theta, \dot{\bm\theta})| c(\bm\theta)=0 \;\text{and}\; \frac{\partial c}{\partial\bm\theta} \dot{\bm\theta}=0 \}$ where 
$\frac{\partial c}{\partial \bm \theta}$ is the Jacobian of the constraints. 
Now we have the new Hamiltonian as
\begin{equation}\label{eq:constrH}
    H(\bm\theta,\bm\phi,\lambda) = \hat H(\bm\theta,\bm\phi) + \lambda^\top c(\bm\theta), \quad \hat H(\bm\theta,\bm\phi) = \psi(\bm \theta) + K(\bm\theta, \bm\phi)
\end{equation}
where $\lambda$ is the Lagrange multiplier.
Then the Hamiltonian dynamics with the above guided Hamiltonian $\hat H(\bm\theta,\bm\phi)$ become \eqref{eq:he-1} and \eqref{eq:he-2} with extra an equation $c(\bm\theta)=0$ and are defined on the \emph{cotangent bundle} $\mathcal{T^*M}=\{(\bm\theta, \bm\phi)| c(\bm\theta)=0 \;\text{and}\; \frac{\partial c}{\partial\bm\theta} \frac{\partial\hat H}{\partial \bm \phi} =0 \}$.
\cite{brubaker12} used a consistent integrator called RATTLE \citep{ANDERSEN_1983,Leimkuhler_1994} to solve the constrained Hamiltonian dynamics in their proposed constrained HMC (CHMC).
\begin{equation}\label{eq:RATTLE}
\begin{aligned}
    \bm \phi^{(\ell+\frac{1}{2})} &= \bm \phi^{(\ell)}-\frac{\epsilon}{2} \left[ \frac{\partial\hat H(\bm \theta^{(\ell)}, \bm \phi^{(\ell+\frac{1}{2})})}{\partial \bm\theta} + \frac{\partial c}{\partial\bm\theta} (\bm\theta^{(\ell)}) \lambda \right] \\
    \bm \theta^{(\ell+1)} &= \bm \theta^{(\ell)} + \frac{\epsilon}{2} \left[ \frac{\partial\hat H(\bm \theta^{(\ell)}, \bm \phi^{(\ell+\frac{1}{2})})}{\partial \bm\phi} + \frac{\partial\hat H(\bm \theta^{(\ell+1)}, \bm \phi^{(\ell+\frac{1}{2})})}{\partial \bm\phi} \right] \\
    0 &= c(\bm\theta^{(\ell+1)}) \\
    \bm \phi^{(\ell+1)} &= \bm \phi^{(\ell+\frac{1}{2})}-\frac{\epsilon}{2} \left[ \frac{\partial\hat H(\bm \theta^{(\ell+1)}, \bm \phi^{(\ell+\frac{1}{2})})}{\partial \bm\theta} + \frac{\partial c}{\partial\bm\theta} (\bm\theta^{(\ell+1)}) \mu \right] \\
    0 &= \frac{\partial c}{\partial\bm\theta} (\bm\theta^{(\ell+1)}) \frac{\partial\hat H(\bm \theta^{(\ell+1)}, \bm \phi^{(\ell+1)})}{\partial \bm\phi}
\end{aligned}
\end{equation}
where $\lambda$ and $\mu$ are Lagrange multipliers associated with the state and momentum constraints (the third and the fifth equations of \eqref{eq:RATTLE}).
This generalizes the leapfrog method to handle the manifold constraints. The proposed state $\bm\theta^*$ is accepted with probability defined by $H$ in \eqref{eq:constrH}.

\subsection{Spherical HMC}
\cite{lan14b} considered HMC defined on a special manifold, hyper-sphere, denoted as $\mathcal S^d=\{\bm\theta\in\mathbb R^{d+1}| \Vert\bm\theta\Vert_2^2=\sum_{i=1}^{d+1}\theta_i^2=1\}$.
This algorithm is particularly useful to handle a class of constraints defined by the following vector ($\bm\beta\in\mathbb R^d$) $q$-norm:
\begin{equation}\label{qnorm}
\Vert \bm\beta\Vert_q =
\begin{cases}
(\sum_{i=1}^d |\beta_i|^q)^{1/q}, & q\in (0,+\infty)\\
\max_{1\leq i\leq d} |\beta_i|, & q=+\infty
\end{cases}
\end{equation}

\begin{figure}[t]
\begin{center}
\includegraphics[width=.9\textwidth, height=.5\textwidth]{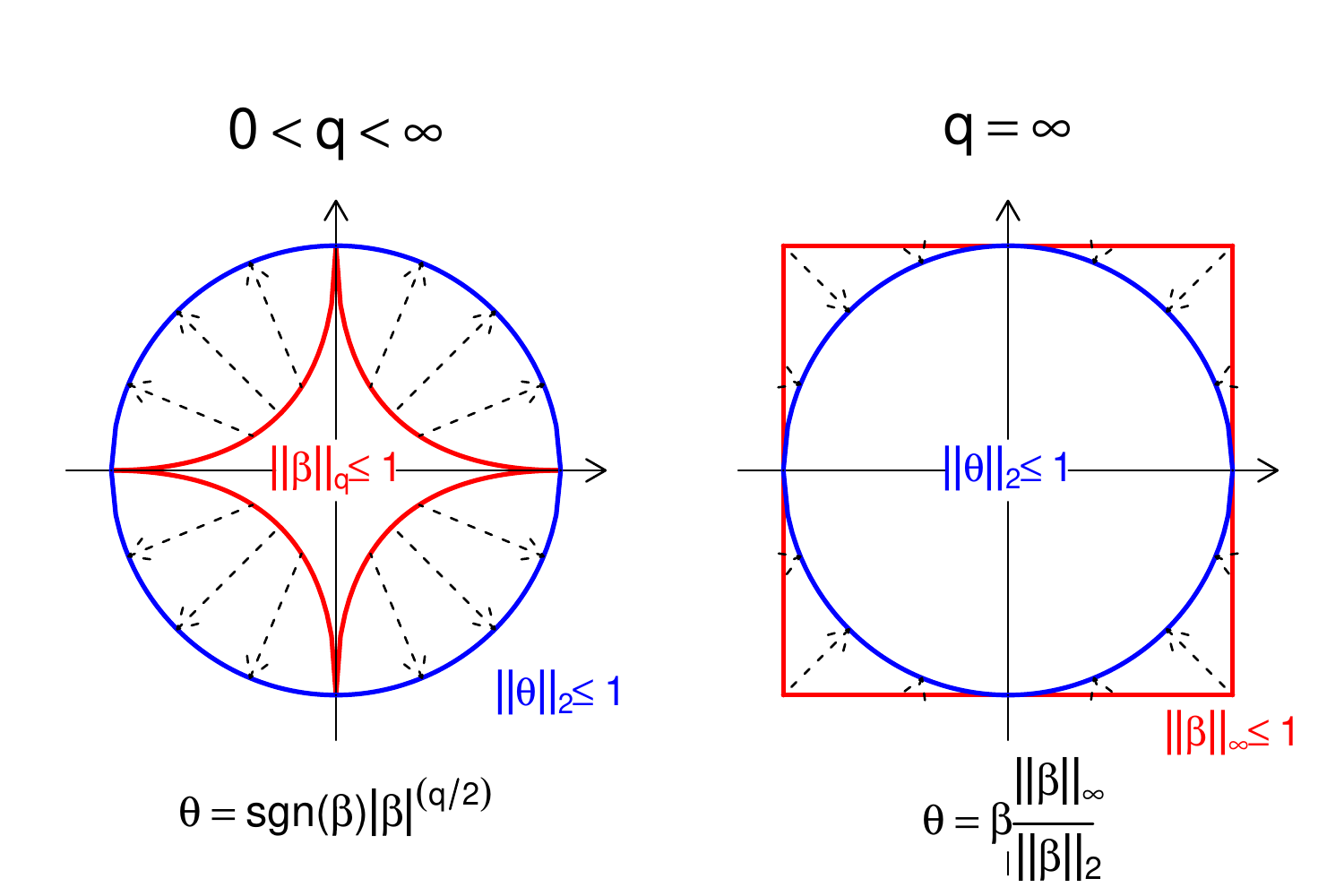}
\vspace{-5pt}
\caption{Transforming $q$-norm constrained domain to the unit ball. Left: from general
$q$-norm domain $\mathcal Q^d$ to unit ball ${\mathcal B}_0^d(1)$; Right: from the unit
cube ${\mathcal C}^d$ to the unit ball ${\mathcal B}_0^d(1)$.}
\vspace{-10pt}
\label{fig:changeofdomain}
\end{center}
\end{figure}

The $q$-norm domain, $\mathcal Q^d:=\{\bm\beta\in\mathbb R^d| \Vert\bm\beta\Vert_q\leq 1\}$, can be transformed to the unit ball $\mathcal B_0^d(1):=\{\bm\theta\in\mathbb R^d| \Vert\bm\theta\Vert_2\leq 1\}$ by either $\beta_i \mapsto \theta_i = \mathrm{sgn}(\beta_i)|\beta_i|^{q/2}$ (shown the left panel of Figure \ref{fig:changeofdomain}) or $\bm\beta \mapsto \bm\theta = \bm\beta \frac{\Vert \bm\beta\Vert_{\infty}}{\Vert \bm\beta\Vert_2}$ (shown in the right panel of Figure \ref{fig:changeofdomain}).

\begin{figure}[t]
\begin{center}
\includegraphics[width=.9\textwidth, height=.5\textwidth]{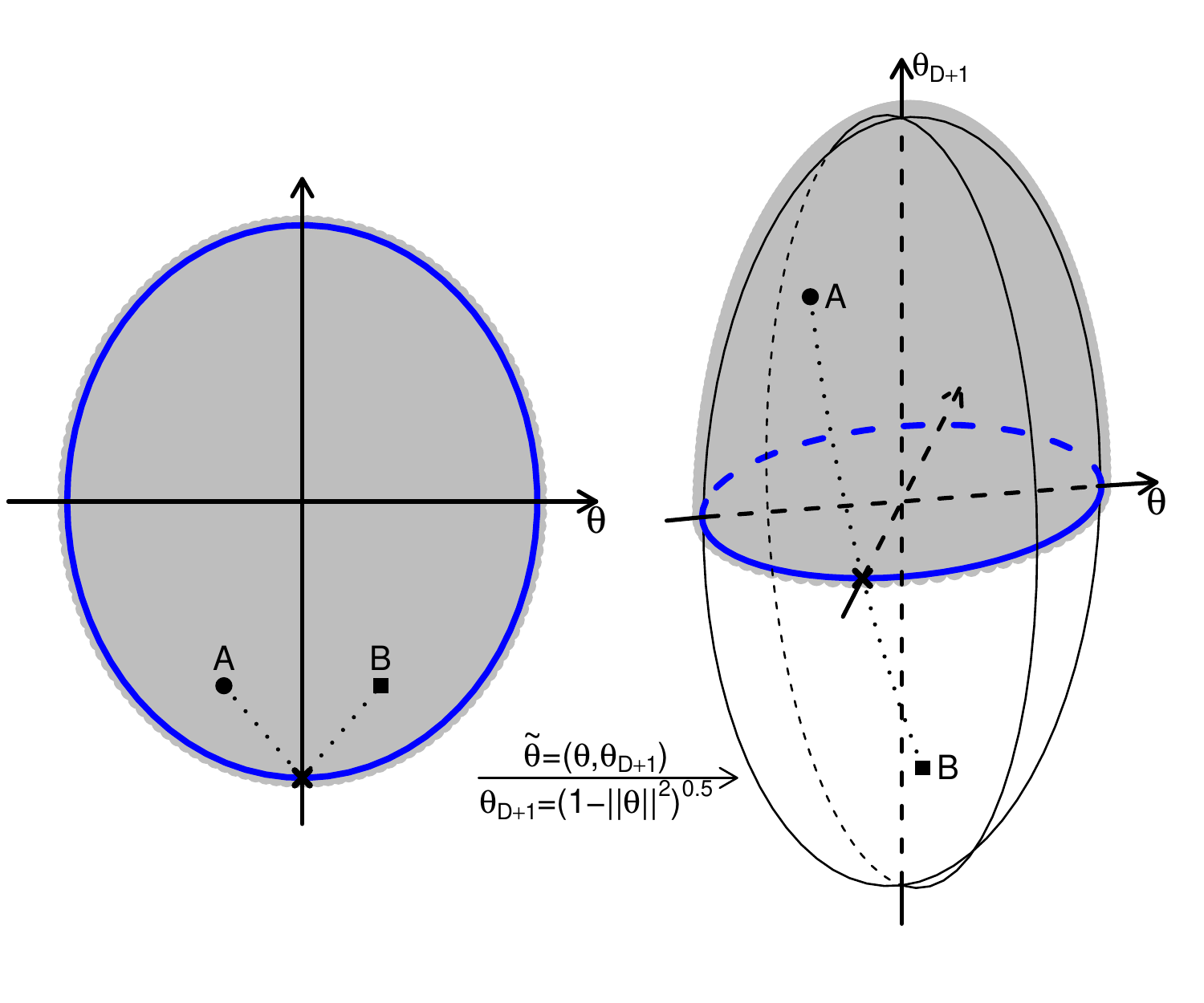}
\vspace{-10pt}
\caption{Transforming the unit ball $\mathcal B_{\bf 0}^d(1)$ to the sphere $\mathcal S^d$.}
\vspace{-10pt}
\label{fig:B2S}
\end{center}
\end{figure}
To define HMC for these constrained distributions, \cite{Lan2016b} proposed an idea of \emph{spherical augmentation} to further map the unit ball $\mathcal B_{\bf 0}^d(1)$ to the hyper-sphere $\mathcal S^d$ by appending an auxiliary variable $\theta_{d+1}$ to the original vector $\bm\theta\in\mathcal B_{\bf 0}^d(1)$ such that the extended parameter $\tilde{\bm\theta}=(\bm\theta,\theta_{d+1})\in \mathcal S^d$.
The lower hemisphere ($\theta_{d+1} = -\sqrt{1-\Vert \bm\theta\Vert_2^2}$) is also identified with the upper hemisphere ($\theta_{d+1} = \sqrt{1-\Vert \bm\theta\Vert_2^2}$) by ignoring the sign of $\theta_{d+1}$.
After collecting samples $\{\tilde{\bm\theta}\}$ using spherical HMC defined on the sphere, $\mathcal S^d$, the last component $\theta_{d+1}$ is discarded and the obtained samples $\{\bm\theta\}$ automatically satisfy the constraint $\Vert \bm\theta\Vert_2\leq 1$.
As illustrated in Figure \ref{fig:B2S}, the boundary of the constraint, i.e., $\Vert\bm\theta\Vert_2=1$, corresponds to the equator on the sphere $\mathcal S^d$. Therefore, as the sampler moves on the sphere, e.g. from $A$ to $B$, passing across the equator from one hemisphere to the other translates to ``bouncing back'' off the boundary in the original parameter space.

On the Riemannian manifold $(\mathcal S^d, {\bf G}(\bm\theta))$ where ${\bf G}(\bm\theta)={\bf I}_d+\bm\theta {\bm\theta}^\top/(1-\Vert\bm\theta\Vert_{2}^2)$ is the \emph{canonical spherical metric}, the tangent space at $\tilde{\bm\theta}$ is defined as $\mathcal{T_{\tilde{\bm\theta}}S}^d=\{\tilde{\bm v}=(\bm v, v_{d+1})\in\mathbb R^{d+1}| \tilde{\bm\theta}^\top \tilde{\bm v}=0 \}$. We have the Hamiltonian \eqref{eq:mhamiltonian} redefined on the tangent bundle $\mathcal{TS}^d=\{(\tilde{\bm\theta}, \tilde{\bm v})| \Vert\tilde{\bm\theta}\Vert_2=1 \;\text{and}\; \tilde{\bm\theta}^\top \tilde{\bm v}=0 \}$:
\begin{equation}\label{eq:HamiltonianSc}
H(\tilde{\bm\theta}, \tilde{\bm v}) = H^*(\tilde{\bm\theta}, \tilde{\bm v}) + \frac{1}{2}\log \det(\bm G(\bm \theta)) , \quad H^*(\tilde{\bm\theta}, \tilde{\bm v}) = U(\tilde{\bm\theta}) + K(\tilde{\bm v})
\end{equation}
where the potential energy $U(\tilde{\bm\theta})=U(\bm\theta)$, and the kinetic energy $K(\tilde{\bm v})=
\frac{1}{2}{\bm v}^\top {\bf G}(\bm\theta){\bm v} = \frac{1}{2}\Vert \tilde{\bm v}\Vert_2^2$ which is defined for the velocity random variable $\tilde{\bm v}\sim \mathcal N(0, \mathcal P(\tilde{\bm\theta}))$ with $\mathcal P(\tilde{\bm\theta})={\bf I}_{d+1} - \tilde{\bm\theta}\tilde{\bm\theta}^\top$ being the projection matrix.

With such guided Hamiltonian function $H^*$ in \eqref{eq:HamiltonianSc}, the Hamiltonian dynamics can be defined on the Riemannian manifold $(\mathcal S^d, {\bf G}(\bm\theta))$ in terms of $(\bm\theta, {\bf p})$, or equivalently as the following Lagrangian dynamics in terms of $(\tilde{\bm\theta}, \tilde{\bm v})$ \citep{lan14a}:
\begin{align}\label{eq:sphLD}
\begin{aligned}
&\dot{\tilde{\bm\theta}} && = && \tilde{\bm v}\\
&\dot{\tilde{\bm v}} && = && -\Vert\tilde{\bm v}\Vert_2^2 \tilde{\bm \theta} - \mathcal P(\tilde{\bm\theta}) \nabla_{\tilde{\bm\theta}} U(\bm\theta)
\end{aligned}
\end{align}
where the projection matrix $\mathcal P(\tilde{\bm\theta})$ maps the directional derivative $\nabla_{\tilde{\bm\theta}} U(\bm\theta)$ onto the tangent space $\mathcal{T_{\tilde{\bm\theta}}S}^d$.
The dynamics \eqref{eq:sphLD} can be split into two dynamics

\noindent
\begin{subequations}
\begin{minipage}{.55\textwidth}
\begin{align}\label{eq:sphLD-U}
\begin{cases}
\begin{aligned}
&\dot{\tilde{\bm\theta}} && = && {\bm 0}\\
&\dot{\tilde{\bm v}} && = && - \mathcal P(\tilde{\bm\theta}) \nabla_{\tilde{\bm\theta}} U(\bm\theta)
\end{aligned}
\end{cases}
\end{align}
\end{minipage}
\begin{minipage}{.45\textwidth}
\begin{align}\label{eq:sphLD-K}
\begin{cases}
\begin{aligned}
&\dot{\tilde{\bm\theta}} && = && \tilde{\bm v}\\
&\dot{\tilde{\bm v}} && = && - \mathcal P(\tilde{\bm\theta}) \nabla_{\tilde{\bm\theta}} U(\bm\theta)
\end{aligned}
\end{cases}
\end{align}
\end{minipage}
\end{subequations}
where the solution to \eqref{eq:sphLD-U} only involves updating velocity and \eqref{eq:sphLD-K} is the geodesic equation on $\mathcal S^d$ with the solution being the big circle.
\emph{Spherical HMC (SphHMC)} proceeds by proposing a joint state by alternate updates according to \eqref{eq:sphLD-U} (lines 5 and 8 in Algorithm \ref{Alg:cSphHMC}) and \eqref{eq:sphLD-K} (lines 6-7 in Algorithm \ref{Alg:cSphHMC}) and accepting or rejecting the proposal based on the acceptance probability defined by $H$.
Algorithm \ref{Alg:cSphHMC} summarizes the details of spherical HMC (SphHMC).
\begin{algorithm}[t]
\caption{Spherical HMC (SphHMC)}
\label{Alg:cSphHMC}
\begin{algorithmic}[1]
\STATE Initialize $\tilde{\bm\theta}^{(0)}$ at current $\tilde{\bm\theta}$.
\STATE Sample a new velocity value $\tilde{\bm v}^{(0)}\sim \mathcal N({\bf 0},{\bf I}_{d+1})$,
and set $\tilde{\bm v}^{(0)} \leftarrow  \mathcal P(\tilde{\bm\theta}^{(0)}) \tilde{\bm v}^{(0)}$.
\STATE Calculate $H(\tilde{\bm\theta}^{(0)},\tilde{\bm v}^{(0)})=U(\bm\theta^{(0)}) + K(\tilde{\bm v}^{(0)})$ 
\FOR{$\ell=0$ to $L-1$}
\STATE $\tilde{\bm v}^{(\ell+\frac{1}{2})} = \tilde{\bm v}^{(\ell)}-\frac{\epsilon}{2} \mathcal P(\tilde{\bm\theta}^{(\ell)}) \nabla_{\tilde{\bm\theta}} U(\bm\theta^{(\ell)})$
\STATE $\tilde{\bm\theta}^{(\ell+1)} = \tilde{\bm\theta}^{(\ell)} \cos(\Vert \tilde{\bm v}^{(\ell+\frac{1}{2})}\Vert \epsilon) + \frac{\tilde{\bm v}^{(\ell+\frac{1}{2})}}{\Vert \tilde{\bm v}^{(\ell+\frac{1}{2})}\Vert} \sin(\Vert \tilde{\bm v}^{(\ell+\frac{1}{2})}\Vert \epsilon)$
\STATE $\tilde{\bm v}^{(\ell+\frac{1}{2})} \leftarrow -\tilde{\bm\theta}^{(\ell)}\Vert \tilde{\bm v}^{(\ell+\frac{1}{2})}\Vert \sin(\Vert \tilde{\bm v}^{(\ell+\frac{1}{2})}\Vert \epsilon) + \tilde{\bm v}^{(\ell+\frac{1}{2})} \cos(\Vert \tilde{\bm v}^{(\ell+\frac{1}{2})}\Vert \epsilon)$

\STATE $\tilde{\bm v}^{(\ell+1)} = \tilde{\bm v}^{(\ell+\frac{1}{2})}-\frac{\epsilon}{2} \mathcal P(\tilde{\bm\theta}^{(\ell+1)}) \nabla_{\tilde{\bm\theta}} U(\bm\theta^{(\ell+1)})$
\ENDFOR
\STATE Calculate $H(\tilde{\bm\theta}^{(L)},\tilde{\bm v}^{(L)})=U(\bm\theta^{(L)}) + K(\tilde{\bm v}^{(L)})$
\STATE Calculate the acceptance probability $\alpha =\min\{1, \exp[-H(\tilde{\bm\theta}^{(L)},\tilde{\bm v}^{(L)})+H(\tilde{\bm\theta}^{(0)},\tilde{\bm v}^{(0)})] \}$
\STATE Accept or reject the proposal according to $\alpha$ for the next state $\tilde{\bm\theta}'$
\end{algorithmic}
\end{algorithm}

\section{Applications of Constrained Sampling}\label{sec:examples}

\subsection{Truncated Multivariate Gaussian}
\begin{figure}[t]
\begin{center}
\includegraphics[width=.9\textwidth,height=0.45\textwidth]{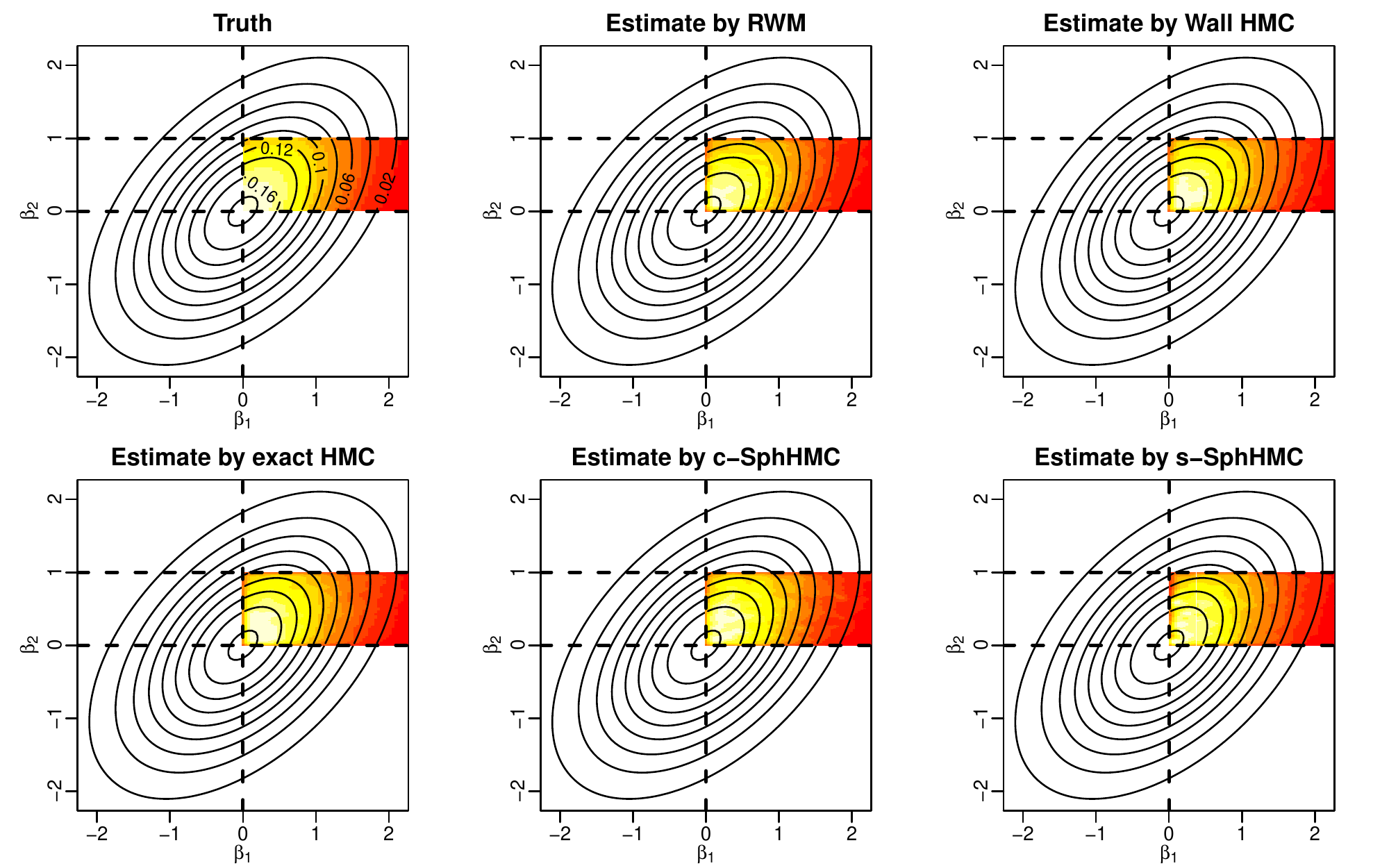}
\caption[Truncated Multivariate Gaussian]{Density plots of a truncated bivariate Gaussian using exact density function (upper leftmost) and MCMC samples from RWM, Wall HMC, exact HMC, c-SphHMC, and s-SphHMC respectively.
Solid elliptical curves always show true unconstrained probability density contours. Dashed lines define linear constrained domains.
Colored heatmaps indicate constrained probability density based on truth or estimation from MCMC samples.}
\label{fig:TMG-sph}
\end{center}
\end{figure}

For illustration purposes, we start with a truncated bivariate Gaussian distribution,
\begin{equation*}
\binom{\beta_1}{\beta_2} \sim \mathcal N\left(\bf{0}, \begin{bmatrix} 1& .5\\ .5 & 1\end{bmatrix} \right), \qquad
0\leq \beta_1\leq 5, \quad 0\leq \beta_2\leq 1
\end{equation*}
This is a box type constraint with the lower and upper limits as ${\bf l}=(0,0)$ and ${\bf u}=(5,1)$ respectively.
The original rectangle domain can be mapped to 2d unit disc $\mathcal B_{\bf 0}^2(1)$ to use c-SphHMC, 
or mapped to 2d rectangle $\mathcal R_{\bf 0}^2$ where s-SphHMC can be directly applied \citep{Lan2016b}.

The upper leftmost panel of Figure \ref{fig:TMG-sph} shows the heatmap based on the exact density function, and the other panels show the corresponding heatmaps based on MCMC samples from RWM, Wall HMC, exact HMC, c-SphHMC, and s-SphHMC respectively. 
All algorithms generate probability density estimates that visually match the true density.
Table \ref{TMG-moments} compares the true mean and covariance of the above
truncated bivariate Gaussian distribution with the point estimates using $2\times 10^5$
($2\times 10^4$ for each of 10 repeated experiments with different random seeds) 
MCMC samples in each method. Overall, all methods estimate the mean and covariance reasonably well.


\begin{table}[ht]
\caption{Comparing the point estimates for the mean and covariance of a bivariate truncated Gaussian distribution using RWM, Wall HMC, exact HMC, and SphHMC.} 
\label{TMG-moments}
\centering
\begin{tabular}{l|c|c}
  \hline
Method & Mean & Covariance \\ 
  \hline
Truth & $\begin{bmatrix}0.7906\\0.4889\end{bmatrix}$ & $\begin{bmatrix}0.3269&0.0172\\0.0172&0.08\end{bmatrix}$ \\ 
   \hline
RWM & $\begin{bmatrix}0.7796\pm0.0088\\0.4889\pm0.0034\end{bmatrix}$ & $\begin{bmatrix}0.3214\pm0.009&0.0158\pm0.001\\0.0158\pm0.001&0.0798\pm5e-04\end{bmatrix}$ \\ 
   \hline
Wall HMC & $\begin{bmatrix}0.7875\pm0.0049\\0.4884\pm8e-04\end{bmatrix}$ & $\begin{bmatrix}0.3242\pm0.0043&0.017\pm0.001\\0.017\pm0.001&0.08\pm3e-04\end{bmatrix}$ \\ 
   \hline
exact HMC & $\begin{bmatrix}0.7909\pm0.0025\\0.4885\pm0.001\end{bmatrix}$ & $\begin{bmatrix}0.3272\pm0.0026&0.0174\pm7e-04\\0.0174\pm7e-04&0.08\pm3e-04\end{bmatrix}$ \\ 
   \hline
SphHMC & $\begin{bmatrix}0.79\pm0.005\\0.4864\pm0.0016\end{bmatrix}$ & $\begin{bmatrix}0.3249\pm0.0045&0.0172\pm0.0012\\0.0172\pm0.0012&0.0801\pm0.001\end{bmatrix}$ \\ 
   \hline
\end{tabular}
\end{table}



To evaluate the efficiency of the above-mentioned methods, we repeat this experiment for higher dimensions, $D=10$, and $D=100$. As before, we set the mean to zero and set the $(i,j)$-th element of the covariance matrix to 
$\Sigma_{ij}=1/(1+|i-j|)$. Further, we impose the following constraints on the parameters,
\begin{equation*}
0\leq \beta_i\leq u_{i}
\end{equation*}
where $u_{i}$ (i.e., the upper bound) is set to 5 when $i=1$; otherwise, it is set to $0.5$.

For each method, we obtain $10^5$ MCMC samples after discarding the initial
$10^4$ samples. We set the tuning parameters of algorithms such that their
overall acceptance rates are within a reasonable range. As shown in Table
\ref{TMG-eff}, Spherical HMC algorithms are substantially more efficient than RWM and Wall
HMC. For RWM, the proposed states are rejected about $95\%$ of times due to
violation of the constraints. On average, Wall HMC bounces off the wall at around 3.81
($L=2$) and 6.19 ($L=5$) times per iteration for $D=10$ and $D=100$ respectively.
Exact HMC is quite efficient for relatively low dimensional truncated Gaussian ($D=10$);
however, it becomes very slow for higher dimensions ($D=100$). In contrast, by augmenting the parameter space, Spherical HMC algorithms handle the constraints more efficiently. Since s-SphHMC is more suited for box-type constraints, it is substantially more efficient than c-SphHMC in this example.

\begin{table}[ht]
\caption{Comparing the efficiency of RWM, Wall HMC, exact HMC, and SphHMC in terms of sampling from truncated Gaussian distributions.} 
\label{TMG-eff}
\begin{center}
\begin{tabular}{l|l|ccccc}
  \hline
Dimension & Method & AP $^a$ & s/iter $^b$ & ESS(min,med,max) $^c$ & Min(ESS)/s $^d$ & speedup \\ 
  \hline
 & RWM & 0.62 & 5.72E-05 & (48,691,736) & 7.58 & 1.00 \\ 
   & Wall HMC & 0.83 & 1.19E-04 & (31904,86275,87311) & 2441.72 & 322.33 \\ 
  D=10 & exact HMC & 1.00 & 7.60E-05 & (1e+05,1e+05,1e+05) & 11960.29 & 1578.87 \\ 
   & SphHMC & 0.82 & 2.53E-04 & (62658,85570,86295) & 2253.32 & 297.46 \\ 
   \hline
 & RWM & 0.81 & 5.45E-04 & (1,4,54) & 0.01 & 1.00 \\ 
   & Wall HMC & 0.74 & 2.23E-03 & (17777,52909,55713) & 72.45 & 5130.21 \\ 
  D=100 & exact HMC & 1.00 & 4.65E-02 & (97963,1e+05,1e+05) & 19.16 & 1356.64 \\ 
   & SphHMC & 0.73 & 3.45E-03 & (55667,68585,72850) & 146.75 & 10390.94 \\ 
   \hline
\end{tabular}
\end{center}
$^a$ acceptance probability\\
$^b$ seconds per iteration\\
$^c$ (minimum, median, maximum) effective sample size\\
$^d$ minimal ESS per second
\end{table}

\subsection{Bayesian Regularized Regression}
\begin{figure}[t]
\begin{center}
\includegraphics[width=.8\textwidth,height=0.4\textwidth]{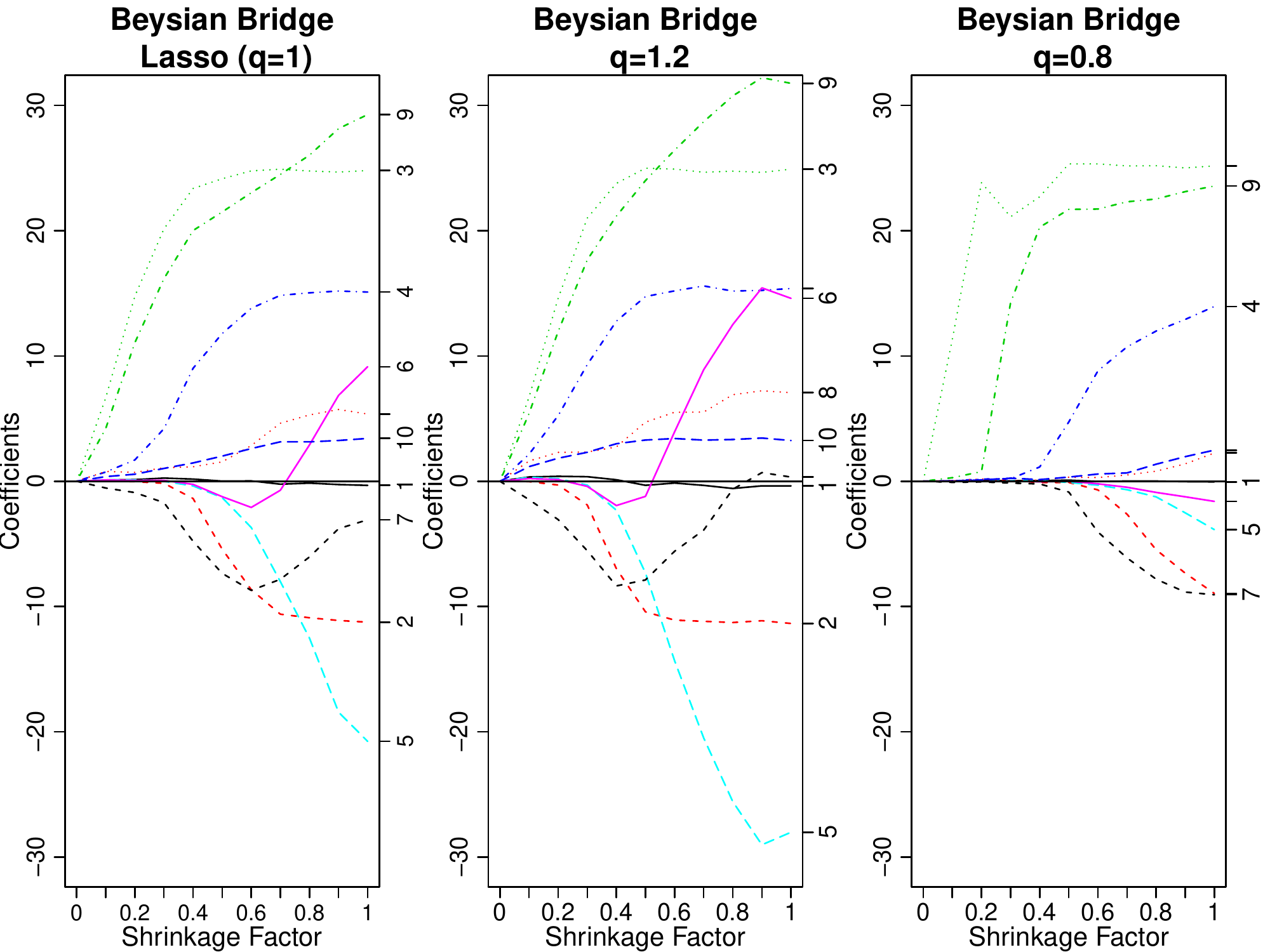}
\end{center}
\caption[Bayesian Bridge Regression by Spherical HMC]{Bayesian Bridge Regression by Spherical HMC: Lasso (q=1, left), q=1.2 (middle), and q=0.8 (right).}
\label{fig:bridge}
\end{figure}

In regression analysis, overly complex models tend to overfit the data. Regularized regression models control complexity by imposing a penalty on model parameters.
\emph{Bridge regression} \citep{frank1993statistical} is a family of regression models where the coefficients are obtained by minimizing the residual sum of squares subject to a constraint on the magnitude of regression coefficients as follows: 
\begin{equation}\label{eq:conLin}
\min_{\bm\beta} \sum_{i=1}^n(y_i - X_i\bm\beta)^2 \quad subject\; to \quad \Vert\bm\beta\Vert_q \leq r
\end{equation}
When $q=1$, this corresponds to \emph{Lasso} (least absolute shrinkage and selection operator) proposed by \cite{tibshirani1996regression} which allows the model to force some of the coefficients to become exactly zero (i.e., become excluded from the model).
When $q=2$, this model is known as \emph{ridge regression}. Bridge regression is more flexible by allowing different $q$ norm constraints for different effects on shrinking the magnitude of parameters (See Figure \ref{fig:bridge}).

\cite{park08} and \cite{hans09} proposed \emph{Bayesian Lasso} by employing a conjugate prior distribution of the form $P(\beta) \propto \exp(-\lambda |\beta|)$. \cite{frank1993statistical} also constructed a complicated prior for the Bayesian inference of bridge regression. With spherical HMC, there is much flexibility in choosing priors with $q$-norm constraints. We can define the following \emph{Bayesian regularized linear regression model}:
\begin{equation}\label{eq:BRLM}
\begin{aligned}
\bm y | \bm X, \bm\beta, \sigma^2_\epsilon &\sim  \mathcal N( \bm X \bm\beta, \sigma^2_\epsilon {\bf I})\\
\bm\beta &\sim p(\bm\beta) \bm 1(\Vert\bm\beta\Vert_q \leq r) 
\end{aligned}
\end{equation}

Figure \ref{fig:bridge} compares the parameter estimates of Bayesian Lasso to the estimates obtained from two Bridge regression models with $q=1.2$ and $q=0.8$ for the diabetes data set (N=442, D=10) studied in \cite{park08}.  
Truncated Gaussian prior $\bm\beta \sim \mathcal N(0, \sigma^2 {\bf I}) \bm 1(\Vert\bm\beta\Vert_q \leq r)$ is considered and posterior samples are collected using spherical HMC algorithm. 
Figure \ref{fig:bridge} illustrates the parameter estimates $\hat{\bm\beta}$ with respect to the shrinkage factor $s:=\Vert \hat{\bm\beta}^{\textrm{Lasso}}\Vert_1/\Vert \hat{\bm\beta}^{\textrm{OLS}}\Vert_1$ varying from 0 to 1, where $\hat{\bm\beta}^{\textrm{OLS}}$ denotes the estimates obtained by ordinary least squares (OLS) regression.
As expected, tighter constraints (e.g., $q=0.8$) would lead to faster shrinkage of regression parameters as we decrease $s$. 
Note, the model \eqref{eq:BRLM} can easily be generalized to generalized linear models or nonparametric models such as Gaussian process regression.

\subsection{Non-parametric Density Estimation}

In this example of non-parametric density estimation, we show how a density function $p(x)$ can be modeled on an infinite dimensional sphere and how the spherical HMC can be applied to the Bayesian inference to efficiently obtain the posterior estimate.

Suppose we want to attribute a smooth density function $p(x)$ to observed data $\{x_n\}_{n=1}^N$ on finite domain $\mathcal{D} \subset \mathbb{R}^d$. Define the space of density functions $p(x)$ and the space of square-root density functions, $q(x)=\sqrt{p(x)}$: 
\begin{eqnarray*}
 \mathcal{P}  := &\left\{ p : \mathcal{D} \to \mathbb R \ |\  p \geq 0, \int_\mathcal{D} p(x)\,\mu(dx) = 1 \right\} \\
 \mathcal{Q}  := & \left\{ q: \mathcal{D} \to \mathbb R \ |\ \int_\mathcal{D} q(x)^2 \, \mu(dx) = 1 \right\}\, ,
\end{eqnarray*}
respectively. Although the space $\mathcal{P}$ contains the functions of interest, we instead opt to deal with the space $\mathcal{Q}$ of square-root densities, which can be viewed as the unit sphere in the infinite-dimensional Hilbert space $L^2(\mathcal{D})$. We model the square-root density with a GP prior multiplied by the Dirac measure restricting the function to the unit sphere:
\begin{align}\label{restrictedGP}
 q(\cdot) \sim \mathcal{GP}(0,\mathcal C) \times \delta(q(\cdot)\in\mathcal{Q}) \, .
\end{align}
where $\mathcal C$ has eigen-pairs $\{\lambda_\ell, \phi_\ell(x)\}$ such that $\mathcal C \phi_\ell(x) = \lambda_\ell \phi_\ell(x)$ for $\ell\in\mathbb N$.

Based on the Karhunen-Lo\`{e}ve representation \citep{wang2008karhunen} of the Gaussian random function, we have the following expansion of $q$:
\begin{eqnarray}\label{klexp}
q(\cdot) = \sum_{\ell=1}^\infty q_\ell\, \phi_i(\cdot), & & q_\ell \stackrel{ind}{\sim} N(0, \lambda^2_\ell)
\end{eqnarray}
If we let $\{\phi_\ell(x)\}$ be an orthonormal basis on $L^(\mathcal D)$, then the unit sphere restriction, $\delta(q\in\mathcal{Q})$, translates to following requirement on the infinite sequence $q:=\{q_\ell\}$:
\begin{equation*}
q \in \mathcal{S}^\infty  =  \left\{ q \in \ell^2 \big|\, \langle q,q \rangle_{\ell^2} = \sum_{i=1}^\infty q_i^2 = 1 \right\} \, .
\end{equation*}
In practice, we truncate the K-L expansion \eqref{klexp} at $L>0$ terms and have $q=\{q_\ell\}_{\ell=1}^L \in \mathcal S^{L-1}$.

Now we have the prior for $q$ and the likelihood of the data $x:=\{x_n\}_{n=1}^N$ given $q$ as follows
\begin{flalign*}
\begin{aligned}
 \pi(q) \propto & \delta(q\in \mathcal S^{L-1}) \prod_{i=1}^L \exp \big( -q_i^2/(2\lambda_i^2) \big)\, , \\
 \pi(x|q)= & \prod_{n=1}^N q^2(x_n) =  \prod_{n=1}^N \left|\sum_{\ell=1}^L q_\ell\, \phi_\ell(x_n)\right|^2 ,
\end{aligned}
\end{flalign*}
We can then apply the spherical HMC to sample from the posterior $\pi(q|x)\propto \pi(q) \pi(x|q)$ which is naturally defined on the sphere $\mathcal S^{L-1}$.

\begin{figure}[t]
    \centering
    \includegraphics[width=1\textwidth, height=.45\textwidth]{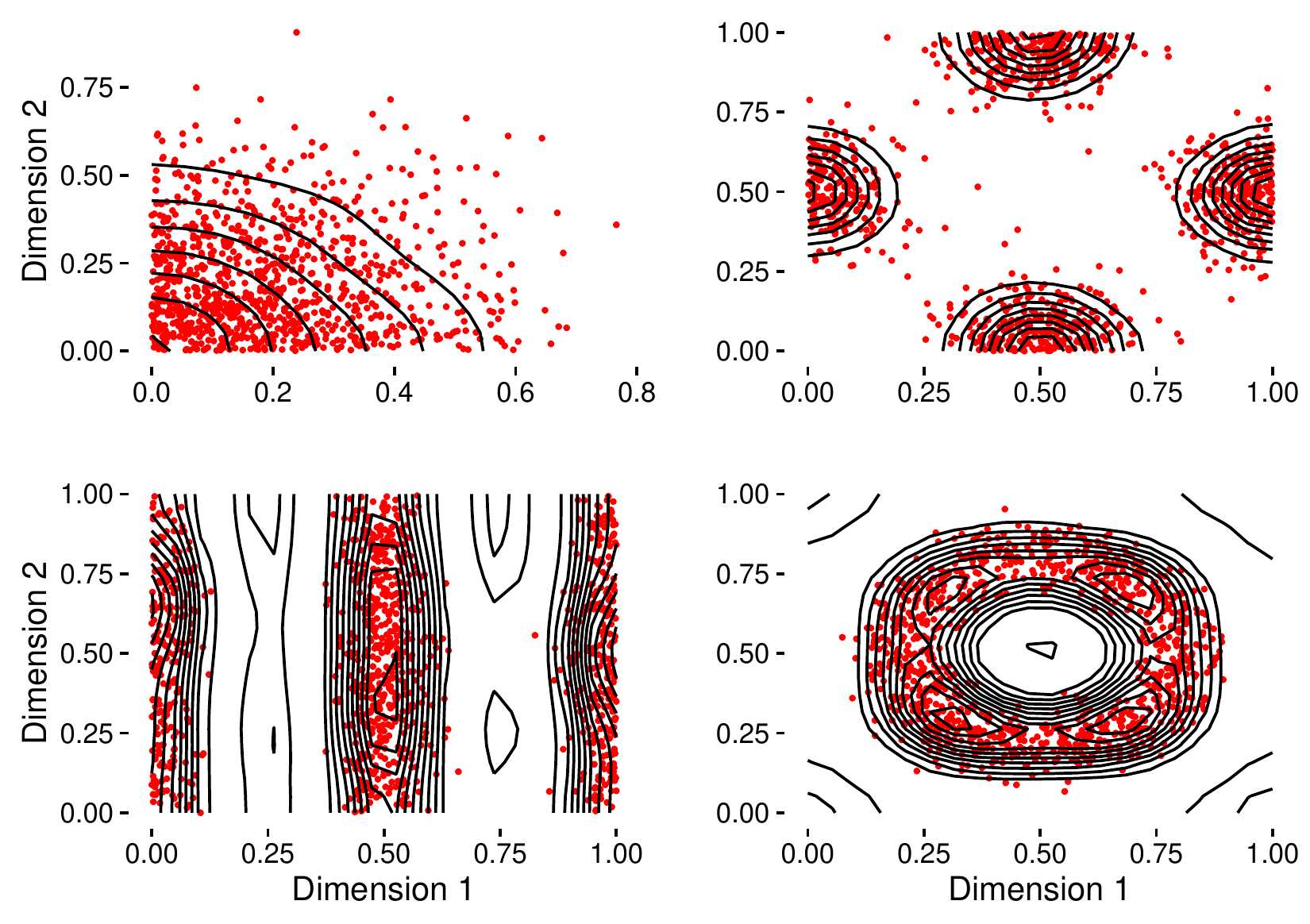}
  \caption{The contours (black) of the posterior median from 1,000 draws of the  $\chi^2$-process density sampler.  Each posterior is conditioned on 1,000 data points (red).}
  \label{fig::2D_plot}
 \end{figure}

Figure \ref{fig::2D_plot} depicts 1,000 data points (red) drawn from four different distributions on the unit square along with the contours of the pointwise median of 1,000 posterior draws from the model $\pi(q|x)$ as defined above. The data in the first three plots are generated using truncated Gaussians and mixtures of truncated Gaussians. The data for the last plot is generated by Gaussian noise added to the uniform distribution on the circle. The model adapts easily to multimodal and patterned data samples. 

\section{Constrained Design of Experiments}\label{sec:doe}

The statistical design of experiments is naturally related to statistical sampling, although they achieve different goals. 
Driven by practical needs, sometimes the design variables have to satisfy certain constraints, which makes the design space irregular, i.e., non-box.

A common type of constrained physical experiment is the mixture experiment, which is widely used in the chemical, pharmaceutical, and food industries. 
The input variables of a mixture experiment are the proportions of different ingredients of the whole mixture. 
Thus the values of all the variables are between 0 and 1 (without scaling or other transformation), and the sum of them must be equal to 1. 
In \cite{piepel2005construction}, the experiment contains 21 ingredients including both box and linear constraints.
Mixture experiments can also have layered mixture structures which makes both design and modeling more complicated \citep{kang2011design,shen2020additive}. 

Many computer experiments also involve irregular constraints, and space-filling design methods have to be adapted to deal with these constraints. 
\cite{stinstra2003constrained} showed a large-scale computer experiment for television tube design that involves 23 input variables and 44 non-box constraints.
\cite{draguljic2012noncollapsing} illustrated a Total Elbow Replacement computer simulation with four input variables and two linear non-box constraints. 

There have been many design methods proposed to handle the constraints. 
\cite{draguljic2012noncollapsing} developed a column-wise construction method for space-filling design. 
In \cite{pratola2017design}, a dense candidate set is generated and then the unfeasible candidate points are removed and the optimal design points are selected from the feasible ones.
Maximin distance and the (robust) IMSPE criteria are used for the space-filling design. 
\cite{huang2021constrained} proposed a constrained minimum energy design method for constructing space-filling designs in any non-regular bounded space, and its key idea is to apply the minimum energy design (a deterministic sampling algorithm) on the target distribution using the probabilistic constraints proposed in sequentially constrained Monte Carlo. 

\begin{figure}
\centering
\includegraphics[width=0.32\textwidth]{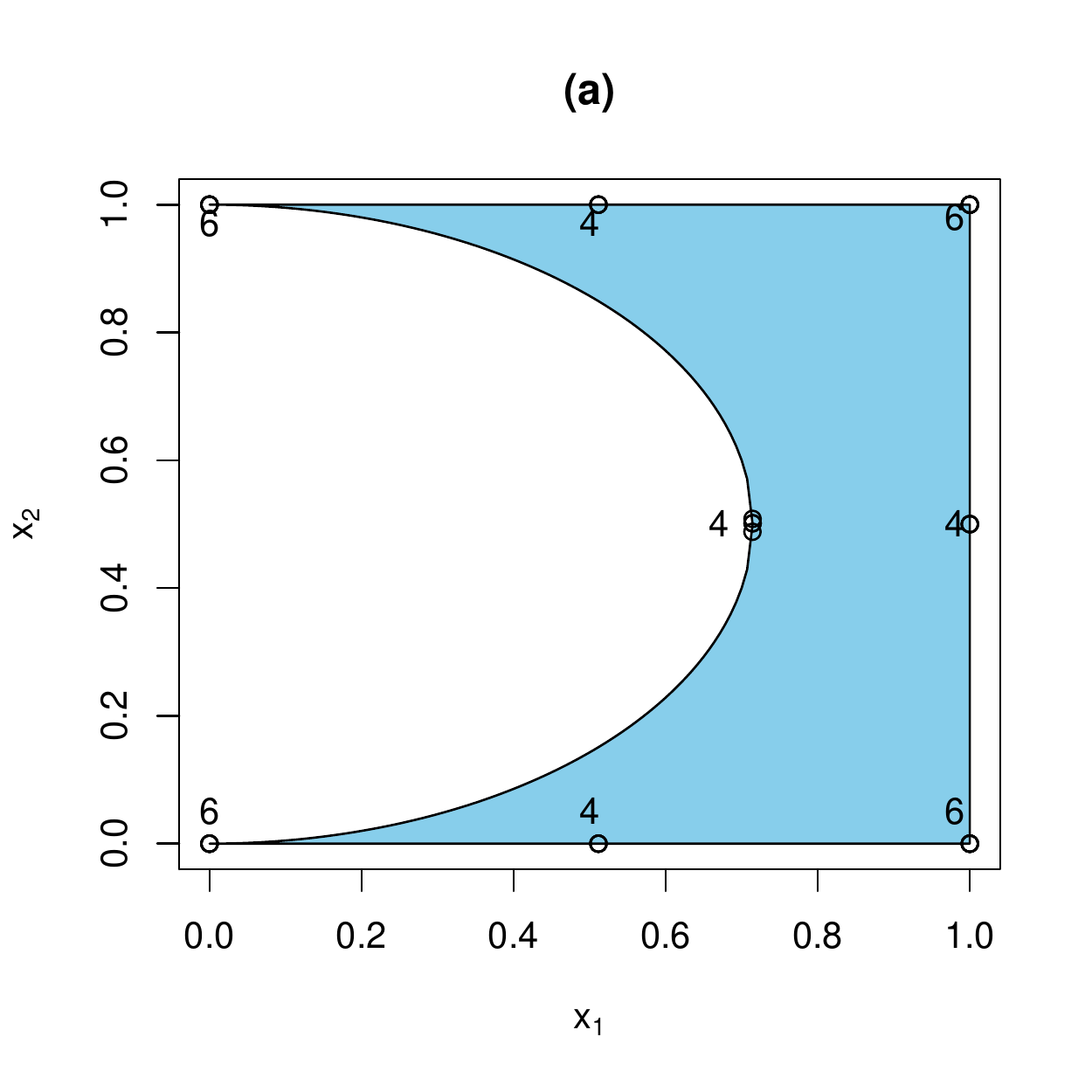} 
\includegraphics[width=0.32\textwidth]{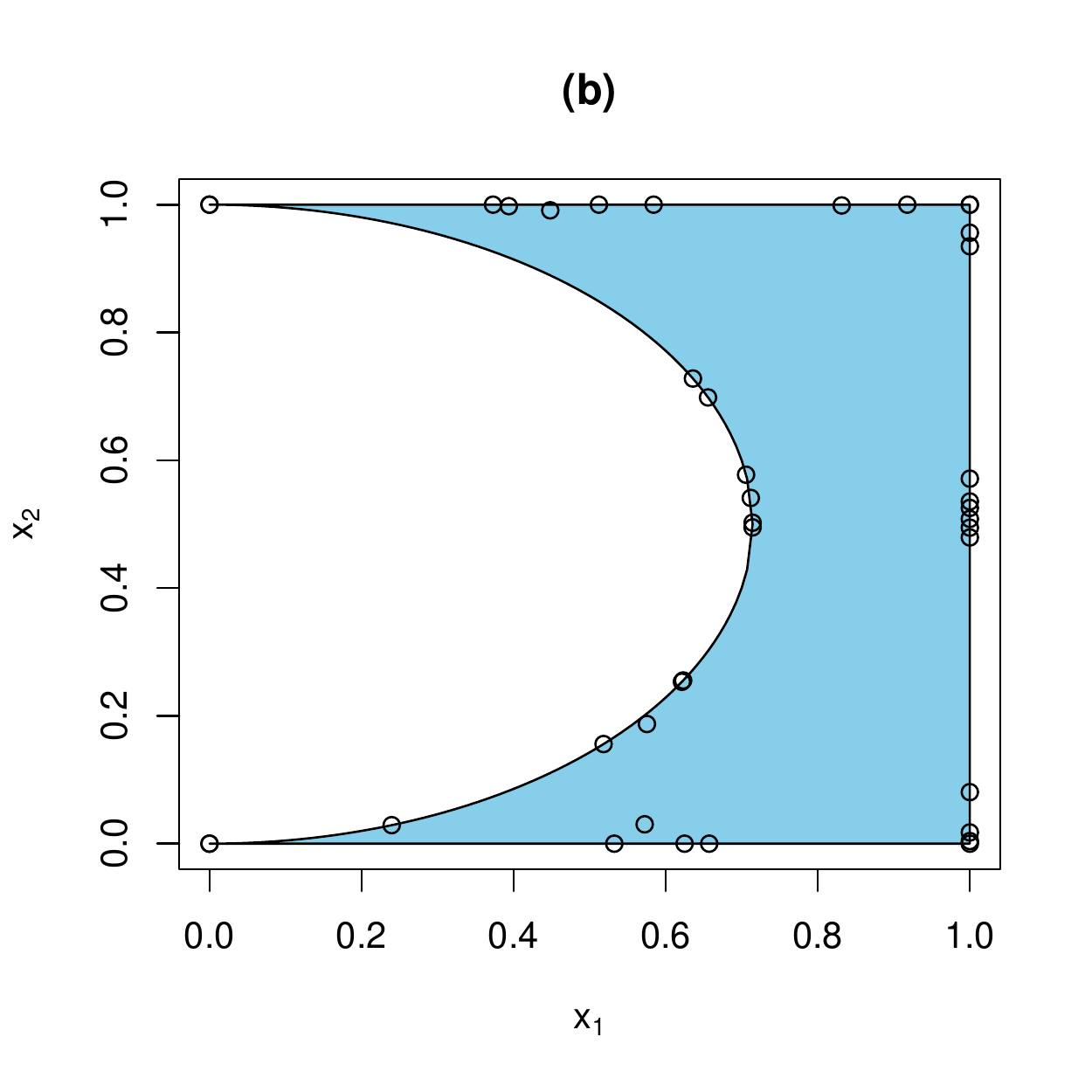} 
\includegraphics[width=0.32\textwidth]{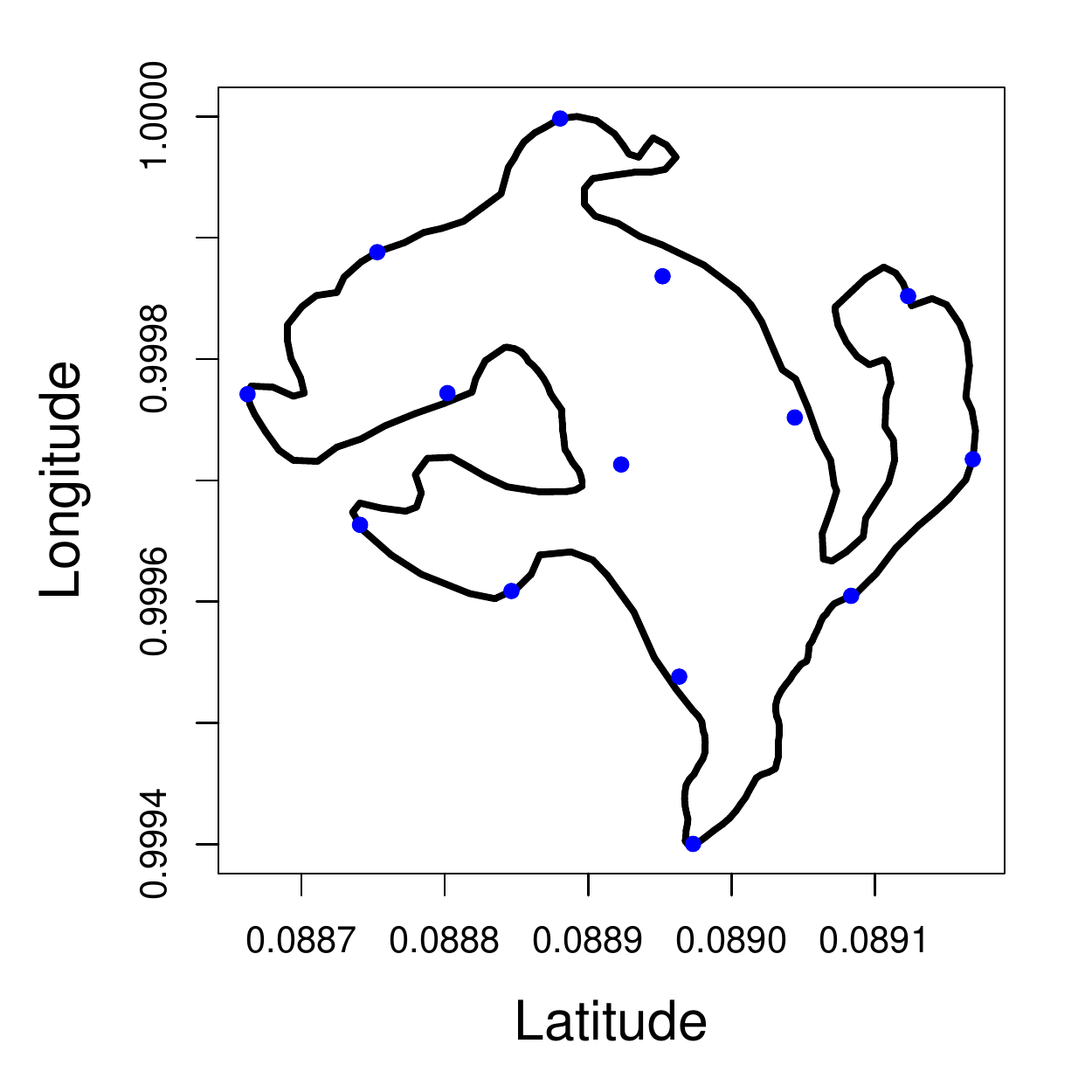} 
\caption{Different constrained designs returned by SCE in \cite{kang2019stochastic}: (a) $D$-optimal design; (b) $I$-optimal design; and (c) space-filling design for the glacier area in the case study of \cite{pratola2017design}.\label{fig:sce}}
\end{figure}

The aforementioned methods are only for constrained space-filling designs, which can be considered as sampling from uniform distributions with constraints. 
\cite{kang2019stochastic} proposed a generic method to construct optimal designs for an irregular constrained design space. 
A stochastic coordinate-exchange (SCE) method is developed. 
In each iteration of coordinate exchange, the multi-dimensional constraints are projected into the dimension of the coordinate to be exchanged (or improved) with the other coordinates fixed at the current values. 
Therefore, the multi-dimensional constraints are reduced to one-dimensional box constraints. 
The generic method can be adapted for different design criteria, including the $D$- and linear-optimal design for physical experiments, and the $\phi_p$-optimal space-filling design for computer experiments. 
Figure \ref{fig:sce} shows three optimal designs constructed by the SCE method for different kinds of constraints. 

\section{Conclusion}\label{sec:end}

In this article, we review in detail the Hamilton Monte Carlo sampling and its variants, and more importantly, how they are modified to overcome various constraints. 
Based on the nature of the algorithms, we categorize them into three groups, rejection, reflection, and reformulation. 
Specifically, we explain three constrained HMC-based sampling algorithms, Wall HMC, Constrained HMC, and Sphere HMC.
Wall HMC is a reflection-type algorithm and the other two belong to the reformulation group. 
Three important applications of constrained sampling algorithms are illustrated. 
They are truncated multivariate Gaussian, Bayesian regularized regression, and non-parametric density estimation. 

Constrained sampling is an important problem and has broad applications in many statistical/machine learning models. 
Some models that are typically not considered to be constrained sampling problems can be solved by constrained sampling algorithms, such as regularized regression/classification and density estimation. 
With the rapid development of AI and big data technologies, future research on this topic is most likely in the direction of creating fast algorithms for high-dimensional distributions with complex constraints, their theoretical foundations, and software implementations. 

\section*{Funding Information}

Shiwei Lan's work is supported by NSF grant DMS-2134256. Lulu Kang's work is supported by NSF grants DMS-1916467 and DMS-2153029. 

\section*{Acknowledgments}

We thank the editor, the associate editor, and reviewers in advance for reviewing this paper. 
We look forward to their comments and suggestions. 

\selectlanguage{english}
\FloatBarrier

\phantomsection
\bibliographystyle{apalike}

\end{document}